# Multiple Scales in the Simulation of Ion Channels and Proteins

March 4, 2019

Bob Eisenberg

Department of Molecular Biophysics and Physiology

Rush University

Chicago IL 60612

*File local reference is*
Multiple Scales in the Simulation of Ion Channels and Proteins SideBar for arXiv.docx

## **Abstract**

Computation of living processes creates great promise for the everyday life of mankind and great challenges for physical scientists. Simulations molecular dynamics have great appeal to biologists as a natural extension of structural biology. Once a biologist sees a structure, she/he wants to see it move. Molecular biology has shown that a small number of atoms, sometimes even one messenger ion, like $Ca^{2+}$, can control biological function on the scale of cells, organs, tissues, and organisms. Enormously concentrated ions—at number densities of ~20 M—in protein channels and enzymes are responsible for many of the characteristics of living systems, just as highly concentrated ions near electrodes are responsible for many of the characteristics of electrochemical systems. Here we confront the reality of the scale differences of ions. We show that the scale differences needed to simulate all the atoms of biological cells are $10^7$ in linear dimension, $10^{21}$ in three dimensions, $10^9$ in resolution, $10^{11}$ in time, and $10^{13}$ in particle number (to deal with concentrations of $Ca^{2+}$). These scales must be dealt with simultaneously if the simulation is to deal with most biological functions. Biological function extends across all of them, all at once in most cases. We suggest a computational approach using explicit multiscale analysis instead of implicit simulation of all scales. The approach is based on an energy variational principle *EnVarA* introduced by Chun Liu to deal with complex fluids. Variational methods deal automatically with multiple interacting components and scales. When an additional component is added to the system, the resulting Euler Lagrange field equations change form automatically—by algebra alone—without additional unknown parameters. Multifaceted interactions are solutions of the resulting equations. We suggest that ionic solutions should be viewed as complex fluids with simple components. Highly concentrated solutions—dominated by interactions of components—are easily computed by *EnVarA*. Successful computation of ions concentrated in special places may be a significant step to understanding the defining characteristics of biological and electrochemical systems. Indeed, computing ions near proteins and nucleic acids may prove as important to molecular biology and chemical technology as computing holes  and electrons has been to our semiconductor and digital technology.

Mark Ratner has been part of at least two enormous revolutions in science. Semiconductor electronics has allowed computer technology to grow by Moore's law[1], giving us pocket computers with more capacity than anyone imagined possible in a room--or in a computer of any size, at any price, in Mark's youth. Molecular biology[2] has allowed us to manipulate the molecules of life with an ease and power equally unimagined in the 1950's.

These two revolutions combined to allow what many view as a new revolution, the computation of proteins in atomic detail. Simulations can be made of the thousands of atoms in a protein, and the tens of thousands of water molecules around it, including a few of the ions in those solutions. Simulations running nanoseconds in full atomic detail are being done all over the world as I write these words (July 2010). The promise is that these simulations can directly compute biological function in atomic detail and thus give us control of biology comparable to our control of semiconductors, with all that implies for medical science and our daily lives.

Molecular dynamics (MD as we will call it) takes the static structures of x-ray crystallography and makes them living objects, reaching towards the real molecules of life in the full reality of their function. Reaching is not grasping, however, as all of us of Mark Ratner's generation were taught in high school. This article is about what is needed to extend the reach of MD so it can grasp the reality of biology. I argue that scaling issues make grasping reality nearly impossible, if MD is done in full atomic detail of real biological systems, as most of them actually function.

Grasping biological function is both easier and harder than grasping physical function. It is easier because there are definite scales for many biological functions. There is no definite *a priori* scale for physical systems but biological systems often have a definite scale—like engineering systems—namely the scale of their inputs and outputs. An amplifier is interesting on the (quite limited) scales that it works. An amplifier is not interesting when light is focused on its input. The time varying potential of light is too fast for it. The function of the amplifier is on a definite time scale. Simulations must deal with rapidly changing voltages, but not so rapidly changing as in light.

Biological systems also work on a definite scale. The output of biological systems occur in seconds and micrometers to meters. Of course, the underlying mechanism span many scales (as we shall see in some detail) and involve a whole range of scales. But, no matter what the scales of the mechanisms, simulations must also calculate the functions of life on the scale that those functions actually occur.

Grasping biological function is harder than grasping some physical functions because so many scales are involved simultaneously in most important biological systems. I argue that scaling issues make grasping biological reality nearly impossible, if MD is done in full atomic detail of biological systems as they actually work and are controlled. I argue that an approach that embraces these multiscale realities will show how MD should be used as one of several indispensable tools in the understanding of biomolecules and their function. Of course, there are exceptional systems that do not require analysis on all these multi-scales, but these are rare and not central to biology as a whole.

The scaling issues facing molecular dynamics involve space, time, concentration and voltage and we go through them one by one (Table 1) using biological function and molecular reality as our guides.

We need to focus on biological function because biological systems are only interesting on the scale (and in the conditions) in which they actually work. Physical systems are interesting on all scales. Engineering and biological systems are not. They are only interesting when they perform their natural functions. They must operate within their design limits or they do not operate at all. With the wrong power supply, amplifiers do not amplify. With the wrong gradients of salt, proteins and ion channels do not conduct. Both engineering and biological systems are robust in one range, but delicate in another. We must compute them both in their functioning robust range. Biological systems should be studied only in their functioning robust range because nothing else is interesting. Outside that range, biological systems are dead

### Table 1

| Computational Scale | Biological Scale | Ratio |
|---|---|---|
| **Time** $10^{-16}$ sec<br>*Vibrations of Bonds* | $10^{-5}$ sec<br>*Action Potential* | $10^{11}$ |
| **Space** $10^{-11}$ m<br>*Side Chains of Protein* | $10^{-4}$ m<br>Large Cell | $10^{7}$ |
| **Volume** | | $10^{21}$ |
| **Spatial Resolution** | | $10^{9}$ |
| **Solute Concentration** | $10^{-11}$ to $2\times10^{1}$ M | $10^{12}$ |

*Scaling restrictions implied by the long range electric field are not clear because the accuracy of the Ewald sum treatment of periodic boundary conditions is not clear. See text*

and of limited interest.

The operating limits of biological systems define the scales on which they must be studied. (Almost all) biological function starts around 100 μsec, reaching to $3 \times 10^9$ sec (~100 years) in fortunate cases. Those are the time scales on which biological function must be studied. Biological structure starts at 10 pm (0.1 Å) and reaches to 10 μm in cells, cm in tissues, and meters in organisms. Those are the length scales on which biological structure must be resolved.

Scales of length involve the size of biological systems in one and three dimensions and the resolution needed to deal with those sizes. The smallest important scale of life is found in its molecules, and particularly in the proteins which do so much of life's work.

**Scaling in Space (one dimension)**. Side chains of proteins control an enormous range of biological function. Changing one side chain can completely alter the function of a protein or ion channel. In some cases, changing one atom can do the trick, just as one (atomic) ion in a channel of a $Ca^{2+}$ sensing protein can switch function entirely. The experimental reality is that structural changes of 1 Å can change biological function on the molecular (nanometer), then cellular (micrometer), tissue (centimeter), and animal (meter) scales. So, simulations in atomic detail must reach from 10 pm (to give decent resolution of one atom) to say 100 μm, if we stop at a representative sample of a nerve fiber and its axon, or much larger if we want to simulate the properties of the real nerve fiber reaching from foot to spinal cord in a human, or elephant. (I use the example of a nerve fiber because its main function is understood from atom, to molecule, to membrane, to cell in considerable detail[3], in the form of theory and computations that a physical scientist would recognize. Nerve function can be understood without much use of arbitrary 'arrow models' with undefined physical basis.) Lengths in one dimension range over 7 orders of magnitude in this realistic example.

**Scaling in Space (three dimensions)**. The scaling requirements of MD in three dimensions are greater. The frightening range of linear scales of $10^7$ become the daunting range of $10^{21}$ if one proceeds without approximation or simplification.

Confronted with length scale ranges of $10^{21}$, it seems obvious that one must try to approximate and simplify. This paper focuses on the underlying problems of the full resolution problem, because so many young scientists assume that is possible. But the general goal, reaching beyond this particular paper, is to motivate, construct, and test multiscale models that use appropriate methods at individual scales and combine those methods in a mathematically defined consistent way. Our goal is to motivate systematic simplifications and approximations to make the problems manageable, and we will discuss how to do that toward the end of the paper.

**Resolution in Space (three dimensions)**. Structures in biology exist in three dimensions and must be resolved in all three dimensions, independent of scale. Resolving a three dimensional structure takes at least 0.1% resolution in each dimension, implying an overall resolution of $10^{-9}$ independent of the particular scale. This resolution is needed to describe a protein well enough to compute its volume, surface area, or the electrical potential around it, if it were a solid macroscopic charged object. The same resolution is needed to reconstruct a cell, tissue or animal. In fact, the difficulties of dealing with three dimensional structures with this resolution are not resolved.

The implications of these resolution requirements are large. Many gigabytes of memory are needed to describe a static three dimensional structure with 0.1% resolution in all directions, with double precision floating point numbers as are required for robust computation. Arrays of this size are difficult to store in memory even today, particularly when various versions are needed for mathematical manipulation. Memory bandwidth does not allow rapid handling of these arrays even in present day computers. Much of the interest of biological systems is in their time evolution. The memory needs for static computation are multiplied by the number of time steps needed to compute time evolution. If the time step is tiny ($10^{-15}$sec), and the time duration is as short as that of the quickest functions of a nerve fiber ($10^{-7}$sec), the dynamic problem is $10^8$ times more demanding than the static one. Simulations of structure reaching to 1 sec are $10^{15}$ times more demanding than the static one.

The static problem itself is demanding. It is not possible yet to solve the partial differential equations of electrostatics in three dimensions with this 0.1% resolution for surfaces as complex as those that define proteins, in any way approaching routine, although the makers of computer games are trying their best, and will surely succeed soon (i.e., within a decade: three or four iterations of Moore's law). The issue is not the complexity of the surface of the protein. For the purposes I have in mind the surface of a rigid protein would need somewhat less resolution than the surface of an animal. The issue is the limitations of Poisson solvers presently available. One imagines that numerical procedures to solve three dimensional partial differential equations with 0.1% resolution exist 'behind the fence(s)'—in weapons laboratories where the nuclear fusion weapons of our nightmares are designed—but that capability is not generally available to outsiders.

**Scaling in Time.** The scaling requirements in time are easy to define for biological simulations. MD simulations must be done with step sizes less than femtoseconds to resolve atomic vibrations. Step sizes of $10^{-16}$ seconds are best but $10^{-15}$ seconds will do. The fastest biological functions (that do not involve light) occur in about $10^{-4}$ seconds. (I have the signals in nerve cells in mind.) There are of course many special properties of proteins that occur in $10^{-5}$ sec or even faster, with proteins involved in photosynthesis and vision being very fast indeed. But the great majority of living processes start around 1 msec and reach as long as $3 \times 10^9$ sec (~100 years).

The gap in time scales between a full resolution treatment of atomic motion and a typical nerve signal is, then, 11 orders of magnitude, $10^{-15}$ to $10^{-4}$. One hundred billion ($10^{11}$) is a very large gap indeed. It corresponds to the gap between a few days on the earth when it was forming $10^9$ years ago, and today. Few would think to compute the properties of the earth today by starting with its properties a billion years ago, computing on a time scale of days for the whole way, the entire time. The reach needed to compute biological function in full detail challenges the imagination—and evades the grasp—of scientists in other fields.

Arguments have sometimes been made that computations on, say, a picosecond time scale can explore 'phase space' and thus deal with biological phenomena on the msec time scale. These arguments have been heard and half-believed by many students and beginning scientists and so I present a counter example here. I hope to make clear the obvious, that if one wishes to study something that takes 1 msec, one must compute on at least a 1 msec time scale.

Imagine a system computed to 100 picoseconds. Imagine another identical system to which a spring, mass, and dashpot are added that create a mechanical resonance that becomes measurable at only 500 picoseconds. All properties computed after 1000 picoseconds will depend dramatically on the resonance. The resonance is not detectable in the short time system. Thus, the short time system cannot reproduce the properties of the resonance. It is obviously possible to make this counter example as realistic and explosive as desired by replacing the resonance with a nonlinear triggered process that can be discontinuously sudden. The conclusion is that computations to a short time will miss long time phenomena. Thus any system must be computed on the time scale on which it functions.

**<u>Scaling in Parameters.</u>** Scaling issues occur in the 'thermodynamic' parameters used to describe life, as well as in time and space. Everyday experience and experimentation show that life involves variables like concentration, average electrical potential, and thus electrochemical potential. The importance of these variables has been known a very long time, by Aristotle and by Galvani and Volta, all of whom were as much biologists (really physiologists) as they were physicists. In a particularly vital example, the heart beat is sensitive to changes in the type and concentration of $Na^+$ , $K^+$, $Ca^{2+}$, and $Cl^-$ ions (among many others). Quite small changes in these concentrations make large changes in function and large changes in concentration are incompatible with normal function: The heart stops.

$Na^+$, $K^+$, $Ca^{2+}$, and $Cl^-$ ions make the plasma needed to sustain the life of cells and proteins. Ions in water are the 'liquid of life', without hyperbole. Anyone who has placed a protein, a tissue or a cell in distilled water has watched the tissue, cell, or protein quickly die or denature. Biological experiments on any scale show that ions in water—***not*** water itself—are the liquid of life. Simulations must then include ions with reasonable realism because ions are needed to keep living things alive, whether the things are proteins, nucleic acids, cells or tissues. Living systems require ions.

Simulating ions in water—as they are necessary for life—is particularly difficult. Most biological systems require mixtures of ions ('Ringer solutions') to exist. Ringer solutions must have $Na^+$, $K^+$, $Cl^-$ and $Ca^{2+}$ each within a certain concentration range. If the ions are absent, or are outside this concentration range, the function of the system is compromised, or in fact the system changes (nearly) irreversibly. Simulations must include realistic concentrations of ions if they are to reproduce experiments. It is not just enough to have one or two ions present. One or two samples of a random variable obviously cannot represent the properties of that variable. That is the entire point of probability theory. 'All' the members of the ensemble must be considered because each member differs from the other. That is what is meant by stochastic. Studying one or two members of that ensemble do not reveal the properties of the ensemble.

One or two ions cannot represent the properties of an ensemble of ions. An ensemble of ions must be simulated if the average properties of an ionic solution are to be computed. Computing the properties of a protein in a Ringer solution, or a mimic of an intracellular solution, requires computation of the (experimentally significant) ions in the solution in the presence of all the others, in realistic concentrations and with statistical reliability.

In fact, gradients of concentration of these ions are the energy sources for an enormous range of cellular signals and processes. The ions have to be present with the right free energy (per mole) in the right place. Ion concentrations are dramatically different outside and inside cells, with "10:1" gradients of $K^+$ and $Na^+$ between the inside and outside of cells, but $10^4:1$ gradients of $Ca^{2+}$. $Ca^{2+}$ is less than $10^{-7}$ M inside cells but (say) $2 \times 10^{-3}$M outside. $Ca^{2+}$ concentration is an important control variable for most proteins that are exposed to the intracellular environment. Variations of a factor of $10\times$ have dramatic often irreversible effects on many of these proteins.

All electrical signaling and a very large fraction of all signaling in cells and tissues are driven by gradients of electrical and chemical potential and not by more 'chemical' processes involving ATP hydrolysis. ATP hydrolysis is used to create these gradients but is rarely used as control signals themselves. A separate set of 'pumps' and transporters is used by biology to maintain gradients of electrochemical potential just as an automobile uses one system (an alternator) to create gradients of electrical potential, and another (battery) to store and allow their use. Gradients of electrochemical potential are used by nearly every cell and organelle in an animal.

In general, MD simulations must deal with thermodynamic variables, including concentration and electrical potentials and ionic currents. Why? Because the concentrations and electrical potentials and ionic currents are the actual function of channel proteins. The channel proteins use concentrations and (average macroscopic) electrical potentials to control (macroscopic) ionic currents that in turn control the electrical signals across nerve and muscle cells, the contraction of cardiac and skeletal muscle, secretion of hormones and an enormous range of biological functions. Concentrations determine the chemical and electrical potentials that are found inside and outside cells. Membranes define cells and separate compartments with different chemical and electrical potentials. Nanovalves called ion channels control the flow of material through the otherwise impermeable membranes that define cells.

Gradients of chemical and electrical potential drive the movement of ions through these nanovalves (nearly picovalves, since the diameter of their charged pore is typically 600 pm, and changes in diameter and charge location of 10 pm are significant) called ion channels. Ion channels are specialized proteins with a hole down their middle that control the flow of ions and electricity through otherwise impermeable membranes. Ion channels have much the same role in living systems that transistors have in engineering systems.[4] Transistors are the fundamental control elements of our digital technology. Ion channels are the fundamental control elements of biology.

Simulations must include the concentrations and conditions in which ion channels work. Simulations must deal realistically with ion channels if they are to be useful. If channels do not function in a particular set of conditions, successful simulations in those literally deadly conditions cannot show them alive. For example, most ion channels 'inactivate' (nearly irreversibly) if the electrical potential across them (the transmembrane potential) is kept near zero. The properties of inactivated channels are difficult to study (if they are inactivated too much) and of limited interest even if they can be studied because they are not functioning the way channels do in real biological situations. MD simulations with zero transmembrane potential must produce inactivated channels if they reproduce the properties of real channel proteins. MD simulations done at equilibrium are likely to have zero transmembrane potential.

Everyday experience and experiments show that these are the variables that biology uses and so these are the variables needed in a direct simulation of ion channel function. The role of electrical potential in nerve conduction and (stimulating) muscle contraction was more obvious to Galvani and Volta than its more physical roles. That is why Galvani and Volta studied nerve muscle preparations. The role of chemical potential (concentration) is just as obvious to every physician. Small changes in $K^+$ concentration, for example, are enough to stop the heart. The concentrations of ions and their free energy per mole (called their 'activity') must be simulated correctly with some precision, as it turns out, because living processes are sensitive to quite small changes in activity of ions.

**Scaling in Concentrations of Ions.** The concentrations of $Na^+$, $K^+$, and $Cl^-$ range across a large scale. Inside and outside cells, concentrations range from millimolar to 500 millmolar. Inside ion channels or active sites of enzymes, however, the concentration of ions is very much larger.

Ion channels and active sites of enzymes typically contain cracks or crevices say 300 $\text{Å}^3$ in diameter lined by amino acids with acidic (negative) or basic (positive) side chains. The concentration—by which I mean number density in molar units—is some 20 molar, compared to the concentration of $H_2O$ in of some 55 molar in distilled water. Nucleic acids are surrounded by narrow regions with enormous densities of ions (typically 10 molar). This enormous density of charge in active sites, channels, and nucleic acids means that the most important locations in proteins are crowded with ions.

Channels and active sites (and the region immediately outside nucleic acids) are very special environments in which the forces of excluded volume and electrostatics are extraordinarily large.

Indeed, any biologist looking at such a special situation on a macroscopic scale would instantly recognize it as an evolutionary adaptation. Just as evolution uses the special properties of certain cells to make a transparent lens, or a rapidly conducting squid axon, so it can use the special properties of crowded ions. The special properties of ions crowded into channels or active sites, or near DNA, can be used to make a nanovalve (ion channel) or a chemical factory (enzyme). The special properties of crowded ions have useful properties that are responsible for the characteristics of channels and enzymes. The special properties of channels and enzymes create biological functions that allow an animal to reproduce more successfully. The special properties of crowded ions are thus a biological adaptation used by evolution. Evolution selects structures that use these properties and makes them an adaptation useful for function.

Biologists start the scientific process of "Guess and Check" with an evolutionary guess. They guess the adaptation and see if that guess leads to useful understanding of function and structure. Biologists think this way for good reason. When they observe unusual structures, they can often guess function, and then design efficient experiments to check that guess. Seeing that a hip joint is a ball and socket leads to an immediate hypothesis about how that joint works, which is far more efficient way to study the joint than writing general mechanical equations for bone.

Most physical scientists are uncomfortable with the idea of adaptation and that is the audience I am writing to, so perhaps I need to be more formal. In my view, unusual adaptations provide productive working hypotheses to investigate, using well defined physical and chemical models, theories and simulations, then checked by direct experiment.

In the "guess and check" of science, good guesses are far more productive than poor ones. Unusual properties of a biological system provide good initial guesses. The crowded ions near DNA, RNA, active sites, and in ion channels should not be ignored. It seems certain to a biologist that evolution has put such special conditions there for a special reason, namely to help the molecules perform their functions. Structural biology provides guesses, experimental biology provides checks. Computational biology is the link between structures and experiments.

(I should add parenthetically that many of the difficulties of doing science arise because the human characteristics of a good guesser are nearly orthogonal to the human characteristics of a good checker. Both wild imagination and compulsive analysis are required to do good science. Both are rarely found in one person. Indeed, one type of person often does not understand the other and finds the other hard to deal with. Guessers and checkers do not always get along.)

Molecular dynamics must then deal with concentration scales from millimolar to many molar, a range of $10^4$. This is the range of concentration of the metal ions $Na^+$, $K^+$, and $Cl^-$ that energize so much of life.

**<u>Scaling in Concentrations of Messengers.</u>** But biology uses ions in another quite different way. It uses some ions as signals, not just as energy sources. $Ca^{2+}$ is used by literally hundreds, probably thousands, of different and distinct signaling pathways in a cell, as a glance at the experimental literature of will quickly show. Thousands of papers are written on signaling molecules every year. Nearly all of these signaling molecules are ions.

The biological systems that use $Ca^{2+}$ as a signal are as different and distinct as the different wires in a computer. And the consequences of cross talk between wires in a computer are replicated in biology. If wires talk to each other loudly in a computer, the computer stops. We say "The computer died" by reflexive analogy to life. If $Ca^{2+}$ signals of different systems are confused in cells, cross talk between otherwise disjoint systems results, and illness and death are a likely consequence. Thus, many biologists study the mechanisms that ensure the integrity of $Ca^{2+}$ signals and that keep different $Ca^{2+}$ signals separate.

Indeed, some of these signaling molecules (particularly $Ca^{2+}$) are necessary for life as well as for function. What I mean is that many intracellular proteins stop functioning—sometimes irreversibly, usually for seconds, minutes, or hours—if they are exposed to unnatural concentrations of $Ca^{2+}$. In general, enzymes, binding proteins, and channels are damaged if their substrate is entirely removed, certainly if there is nothing to replace the substrate in the binding sites. Many, even most channel proteins, 'die', in the sense that they drastically (and more or less irreversibly) change properties if they do not have the right mixture of chemicals surrounding them or if they are not maintained with the right electrical potential across them.

In fact, there are hundreds of hormones, vitamins, messengers, and other organic ions that control the function of proteins, enzymes and ion channels. These go by many names: they were called enzyme co-factors in Mark Ratner's undergraduate years as summarized in the classic tome[5]. The important point is not what they are called. The important point is that the concentration of these signaling molecules controls biological function and the concentrations of these signaling molecules are small, ranging from $10^{-7}M$ to

$10^{-11}$ M.

Finding the right conditions to ensure survival of (function in) proteins is the art of much experimental biology, from microbiology (growing bacteria), to immunology (where it is particularly hard to establish reproducible conditions for various immune responses), to enzymology, to channel biophysics.

Simulations must establish the same conditions for survival as experiments, if one wishes to simulate living functioning proteins. Simulations if successful must reproduce the phenomena of life. Thus, simulations must reproduce the essential conditions of life. If experiments require less than $10^{-6}$ M calcium to maintain function, then simulations must contain less than $10^{-6}$ M calcium. If a channel requires a maintained electrical potential close to -90 mV across it, the simulation must maintain that potential. Simulations of protein folding are likely to be confusing if they do not include the ions needed to allow normal folding in an experimental system. If conditions are unphysiological, the living system will die (i.e., change irreversibly into another system that does not function).

Simulations of channel proteins are likely to give strange results if they do not maintain a resting potential. Maintaining a resting potential across a channel protein is a challenge in simulations using equilibrium assumptions that preclude flow: in real biological systems membrane potentials are nearly always accompanied by flow. In biological jargon[6], not all permeable ions have the same reversal (i.e., equilibrium) potential. Most real proteins require a maintained potential. Most real proteins 'inactivate' more or less irreversibly to a nonfunctional state if there is no electrical potential across them, as we have mentioned.

What are the implications for scaling? The scaling requirements required to deal with trace concentrations of controlling molecules are severe. Concentrations of $10^{-7}$ M $Ca^{2+}$ must be simulated for many systems. Concentrations of $10^{-11}$ M must be simulated to deal with a range of hormones.

Simulating concentration this small requires a staggering number of water molecules. For example, if a simulation needs $10^3$ $Ca^{2+}$ to have a decent estimate of concentration, then a simulation needs 55 moles of water for every $10^{-7}$ moles of $Ca^{2+}$, meaning one needs $6 \times 10^{11}$ molecules in the simulation, along with the $10^3$ $Ca^{2+}$ ions. Simulations of proteins that depend on $Ca^{2+}$ as a signaling molecule typically need $2 \times 10^{13}$ if all atoms of water are included. Hormones might need as much as $2 \times 10^{17}$ atoms.

**Scaling for Electrical Potential.** Scaling requirements for the electrical potential are hard to specify since the electric field is both short and long range. The electric field that controls nerve function, for example, extends millimeters in vertebrate nerve. There can be no ambiguity about this experimental reality. The electrical potential at one location in a nerve fiber controls the function of individual channels in the membrane of the nerve millimeters away. Measurements of a single channel in a nerve fiber demonstrate this experimental fact, as verified everyday by laboratories doing patch clamp experiments.[7]

The electric field of nerve cells is not screened in a Debye length any more than the electrical signal in a telegraph cable under the sea is screened. The assumptions used in the calculation of screening in equilibrium ionic systems do not apply. 'Sum rules'[8] for infinite equilibrium systems without charge on their boundaries do not apply to membrane potentials of cells because cells are finite size nonequilibrium systems with significant charge on their boundaries. Life and experiments occur in finite size systems and living systems and experimenters go to enormous lengths to control the properties of the boundary of these systems. Theorems that ignore boundary conditions may not apply to experiments and living systems with finite boundaries.

Systems without important boundaries occur in biology. The bulk solutions outside cells do not have important boundaries, for the most part. Electrical potentials in bulk solutions also spread very long distances on short time scales, before the screening phenomena of the sum rules comes into play. Sum rules and screening typically take tens of picoseconds to develop. Before then, the spread of potential is more or less that in a dielectric, and details of the shape and type of boundary conditions very far from an atom are important.

The time scales on which screening develops are easily measured experimentally. These are the time scales that determine the linear electrical properties of an ionic solution, its conductance and admittance (in the language of electrical engineering), and its conductance and dielectric coefficient in the language of classical electrochemistry. Many volumes of such measurements have been published, for example, for ions in water[9,10] in many concentrations of many types of ions. Fewer measurements of properties of mixtures are reported because they are so hard to interpret, I suspect.

There is little discussion of the time course of screening in the MD literature. Most of the calculations of MD are executed in a time scale of femtoseconds, far faster than picoseconds. Thus, at the times involved in every MD calculation, electric fields spread very far; at long times, achieved only recently (i.e., the last decade) in MD simulations, electric potentials spread a tiny distance.

At long times ionic solutions are screened, and potentials spread only a few Debye lengths (say 1 nm in typical biological extracellular solutions). The spatial resolution needed in simulations of ionic solution then is very different at short and long times. At short times the spatial resolution needed is very coarse but the spatial domain is macroscopic and must include boundary conditions on the electric field. At long times, the spatial resolution needed is very fine but the spatial domain is small and does not need the boundary conditions required at short times. It is interesting that numerical methods used in simulations of MD or Brownian dynamics have not taken advantage of these screening properties of electrolytes, as far as I know.

One would imagine an integrator could be constructed with coarse spatial resolution at short times and fine spatial resolution at long times. Combining the two methods would, of course, be a problem, but the fact that the long time integrator would not depend on far (spatial) field boundary conditions might be a great help. A natural multiscale integrator and treatment might result.

Most MD simulations involve short times, before screening is established. The electric field involves macroscopic numbers $10^{23}$ of ions. The electric field also involves macroscopic numbers whenever it is involved in functions of life in nerve and muscle cells, for example, because those functions are known experimentally to extend over macroscopic distances, even meters in extreme cases, like the motor or sensory nerves of large mammals.

These macroscopic effects of the electric field are customarily handled in MD by periodic boundary conditions implemented with Ewald sums of various types that are supposed to compute the macroscopic electric field correctly even though they compute in an atomic scale domain (involving say 50,000 atoms and say cubes 100 Å on a side). These procedures are difficult to extend to nonequilibrium situations where gradients of electrical potential are important. Nonequilibrium systems have flows. Equilibrium systems calculated in most MD simulations do not have flows. Flows are found in channels, membrane transporters, and membrane proteins, in nearly all cells, under nearly every natural condition.

Nonequilibrium conditions cannot be easily finessed. Flows are directly involved in a wide range of biological function. Ion channels almost always work away from equilibrium. Transporters and pumps are far from equilibrium. Many enzymes work away from equilibrium. If MD wishes to simulate ion channels, transporters, pumps, many enzymes and living systems, it must include flows of ions and charge, and so it must be extended to nonequilibrium systems. Extension of equilibrium analysis to near equilibrium by a Green-Kubo type treatment is not very helpful if nonlinear behavior is used to create a device with properties distinct from an equilibrium linearized system. That is the reason Green-Kubo treatments of transistors are not prominent in the semiconductor literature. Transporters, pumps, and channels with significant coupling behavior between fluxes (e.g., 'single file' channels) are likely to be too far from equilibrium to allow simple analysis. Simple channels with nearly linear IV relations might be better targets for this approach. However, most channels show quite nonlinear IV relations in some sets of ionic solutions and those are often the solutions most useful in solving the inverse problem, namely in measuring the distribution of fixed charge in a channel.[11]

**<u>Nonequilibrium simulations in computational electronics</u>**. Semiconductor physics and computational electronics[12-14] have studied nonequilibrium situations in particle simulations for a very long time. Semiconductor physics and computational electronics have simulated swarms of holes and electrons, using entirely classical approaches, in which quantum mechanics does not appear at all, since the 1980s.

It is striking that periodic boundary conditions are never used in the calculations of computational electronics[12]. The reason is clear. Devices cannot be (spatially) periodic systems if they have inputs and outputs. The essence of a device is its distinct inputs and outputs. The potential is not the same at the input and output (nor is the current flow). Spatially nonuniform boundary conditions are needed to describe devices. Semiconductor simulations are designed to deal correctly with inputs and outputs, to be sure the boundary conditions are always simulated correctly—because those boundary conditions are the essence of a device, its inputs and outputs and power supply—even when the simulations are done with swarms of interacting particles.

The periodic boundary conditions used in MD may or may not adequately represent the electric field over long ranges in equilibrium systems. I cannot tell because I cannot find simple checks of Gauss' law that analyze these conditions. The semiconductor community checks its computation of the electric field by verifying Gauss' law on a variety of scales, some comparable to the particle size, some much larger. Gauss' law is checked with surfaces that are not parallel to the natural surfaces of the system or to surfaces assumed in periodic boundary conditions. It would be comforting if the various Ewald sum methods were shown to satisfy Gauss' law on scales comparable to atoms, on scales comparable to the period assumed in the periodic boundary conditions, and on scales much larger than that period.

These uncertainties in the treatment of the electric field in MD are large and so I will not consider problems of scale arising from the electric field further: I do not want to speculate in an argumentative way. I confine my scaling arguments to simple cases where it is clear what is involved.

**Summary of Biological Scales.** Aside from the electric field, we are thus confronted with scale issues of $10^7$ in linear dimension, $10^{21}$ in three dimensions, $10^9$ in resolution, $10^{11}$ in time, and $10^{13}$ in particle number (to deal with concentrations of $Ca^{2+}$).

**All scales appear at once**. These many and different scales occur all at once in functioning biological systems. Indeed, typical proteins, channels and nucleic acids involve all these scales in their typical function. Thus MD simulations must be able to deal with all these scales at once. This seems a daunting problem, probably one that cannot be solved. If one imagines the computational issues produced by interactions among this many particles with this spatial resolution on these spatial scales over this duration of time, one is chastened. These seem in principle uncomputable, if long range forces are involved. Electric fields are long range, and nearly all biomolecules bristle with charge[15] that can produce long range electric fields. In that case, one must deal with numbers of calculations beyond astronomical, involving the factorial of the number of particles over these spatial and temporal scales. It is natural that simulations have tried to do parts of a problem, dealing with pieces of the biological situation, doing what they can, hoping to find some way or other to deal with the other pieces, and with the totality of the scaling and sampling problems

of MD

In my view, MD has reached the point where an explicit multiscale analysis is needed. Certainly, if simulations are to confront biological reality, they must deal with the scales found in the real systems. MD simulations in full atomic detail of biological function are not likely to succeed quantitatively until they are embedded in multiscale analysis in my view.

This is not to argue that MD simulations of reduced systems and reduced complexity are not valuable. Indeed, the motivation for this paper is exactly the opposite. I argue that MD is an irreplaceable—extraordinarily important—tool when used properly.

**Role of Molecular Dynamics**. MD is properly used as an extension of structural biology in my opinion. MD shows us how structures move and which motions are important. MD is an essential tool for dealing with the reality of biological structure and the need to reduce complexity in our models. MD can help us guess biological adaptation intelligently. MD can tell us what to leave out in reduced models and what to focus on as we try to make reduced models of biological function.

MD is an essential component of a multiscale approach to computing biological function but it is only one part of that approach.

**What else is needed, beyond Molecular Dynamics?** Some needs are clear and definite in my opinion and some are still vague, a matter of investigation.

What is clear is that we must include the thermodynamic variables concentration and electrical, and chemical potentials of ions with reasonable accuracy because experiments require that accuracy. Biological function in fact depends on and is controlled by concentration, potential and chemical potential with some sensitivity. Even the names of ion channels—sodium, potassium, calcium channels—cannot be determined if the chemical potentials of these ions are unknown. Experiments identify and name channels by comparing chemical potentials with experimentally determined 'reversal potentials', the electrical potential at which the current measured in a channel reverses direction. Simulations must be checked and calibrated to be sure that they give estimates of chemical potentials that are sufficiently accurate for this purpose. Reversal potentials must be measured within a few millivolts to evaluate channels properly in the laboratory; chemical potentials must be computed with similar accuracy, i.e., to better than $0.1(k_B T / e)$, or ~2.5 mV.

It seems clear that simulations must be carefully calibrated to be useful. Work in this direction is just beginning (see reference[16] and the literature cited there). A great deal of attention will be needed to calibrate simulations if they are to deal with experimental and biological reality, because the calibration must be done over a range of concentrations in solutions that are mixtures of many ions, including divalent $Ca^{2+}$.

There are many physical and chemical issues involved in such calibrations and this is not the place to engage in prolonged speculation concerning the difficulties. There is a general problem that needs mentioning however, because it seems finally on the way to resolution, after plaguing biophysics and physical chemistry since their beginnings in the 19th century.

**Biology occurs in mixtures of ions**. Sydney Ringer discovered that the heart, and then muscle, and all cells, require a specific mixture of ions, chiefly $Na^+$, $K^+$, $Ca^{2+}$ and $Cl^-$ if they are to survive. Mixtures of ions are particularly hard to calibrate. Physical chemists have shown in innumerable experiments that the simplest properties of mixtures of ion solutions (i.e., the 'colligative' properties of density, freezing point depression, and boiling point elevation) along with all more subtle properties (mobility, conductance, free energy per mole, called activity) depend on the interactions of all ions when solutions are reasonably concentrated, say beyond 20 mM[10,17,18]. It is crucial to understand that the properties of individual ions depend on the concentrations of every other type of ions in these solutions. That is why they are nonideal solutions, not approximated at all by the properties of ideal gases from which the science of thermodynamics and statistical mechanics grew. Indeed, scientists who must be able to predict the properties of mixtures of ionic solutions use descriptions of enormous complexity, equations of state involving tens sometimes hundreds of parameters[19-21]. Mixtures of ions are of such importance that these parameters are measured experimentally and these unwieldy expressions are used in design by chemical engineers every day.

Ionic solutions have usually been treated as simple fluids [22-24] with complex properties [10,17,18,25] and the enormous literature of ionic solutions and mixtures has been cast in that mold. I hope I insult no one when I say that theory has been less successful than its authors would wish. Solutions made from one salt (e.g., one cation and one anion, like NaCl) can be dealt with some success at equilibrium. More complex solutions, made of mixtures of ions like $Na^+$ and $Cl^-$ and $Ca^{2+}$ and (two) $Cl^-$ are a serious challenge to simple theories even at equilibrium when one wants only to know the free energy per mole, or the freezing and boiling points and vapor pressure. Simple theories of nonequilibrium properties like conductance are a challenge for one salt say $Na^+$ $Cl^-$ [26] Few theories even try to deal with the nonequilibirum properties of mixtures like NaCl mixed with $CaCl_2$ in water.

I believe the reason for these difficulties is that those theories treat ionic solutions as simple fluids in which (in the ideal case) there are no interactions. But interactions dominate the properties of ionic solutions. Speaking crudely, everything interacts with everything else. The properties of every ion are affected by all the other ions, not just other ions of the same type. Interaction terms have to be added into theories of simple fluids, by hand, and the resulting expressions and parameters are multifaceted in their complexity.

I believe we should take a different approach. We should view ionic solutions as complex fluids with simple components, not as simple fluids at all. Complex fluids are fluids in which everything interacts with everything else. We need a mathematics that handles interactions in general, and then simplifies them to the special cases of biological interest. I will argue that the variational calculus, specifically the energetic variational approach *EnVarA* developed by Chun Liu [27-30] provides much of what we need to deal with ions as complex fluids.

**Ions as Complex Fluids**. We should view ionic solutions as complex fluids because ions come 'in pairs'; that is to say, electrostatic interactions are so strong that ions come (always) in (strictly) neutral combinations. The interactions between positive ions (cations) and negative ions (anions) are so strong that deviations from electroneutrality are always tiny. Strong deviations would produce electric fields comparable to the electric field between valence electrons and nuclei inside an atom. Such strong electric fields would destroy these atoms, producing atomic plasmas incompatible with life.

The salts which dissolve in water to create ionic solutions are always strictly neutral[31]. If the salts are made of ions with equal charge (i.e., valence) like $Na^+$ $Cl^-$, ions come in pairs; the neutral combination (which is in fact the definition of a 'molecule' in the periodic lattice of a salt crystal) has two atoms. If the salts are made of elements with unequal charge (like $Ca^{2+}$ $Cl_2^-$), the neutral molecule has three atoms. The macromolecules of life—proteins, nucleic acids, and lipids—always appear in electroneutral combinations with ions (and/or each other). The 'permanent' charges of DNA and proteins (that chemists call their acid and base groups) are balanced by an exactly equal number of ions. Molecular biology is the science of complex fluids with complex elements. The fluids of life are mixtures of ionic solutions, and the macromolecular minielements called organic molecules (like glucose or amino acids), proteins, nucleic acids, and lipids.

In biological and chemical solutions the amount of positive and negative charge in a volume are nearly the same. Macroscopic and mesoscopic amounts of ionic solutions (e.g., tissues and cells in the biological context) are equal within a tolerance of the order of $10^{-15}$. The 'sum rules' of equilibrium statistical mechanics[8] are an expression of the enormous strength of electrostatic interactions that enforce electroneutrality. Even atomic scale systems, active sites of enzymes or pores in channel proteins where dimensions of 1Å are significant, have deviations much less than $10^{-3}$.

Statistical mechanics arose from the treatment of ideal gases of uncharged particles that hardly interact. Statistical mechanics has been extended to deal with simple fluids with great success even when they are nonideal[22-24]. These nonideal fluids have significant hard core interactions caused by the finite volume of molecules that do not overlap. Statistical mechanics has been less successful in dealing with the experimental properties of ionic solutions [10,17,18,32]. Theories of even the fundamental property of solutions (the free energy per mole of each component) have not been particularly successful (see [32] for references) even in solutions of one salt (e.g., NaCl in water). In mixed solutions, like those of living systems, success is even more limited and descriptions used in technological applications (which have to get their predictions right!) often involve large numbers of empirical parameters.[19-21,33]

Molecular dynamics simulations have not escaped these difficulties. These difficulties are not restricted to macroscopic 'mean field' type models. Molecular dynamics uses force fields that are nearly always calibrated under ideal conditions of zero concentration. The force fields of molecular dynamics are not designed to deal with finite concentrations of ions, or mixtures of different types of ions because they are not designed to deal with three body (or n body) problems. It is a matter of mathematics that two body forces cannot uniformly approximate three body interactions; indeed they cannot approximate three body interactions over a wide range of conditions or concentrations, as occur in biology. When atomic scale simulations are used to compute macroscopic systems, they must be calibrated[34] to show that they compute properties actually measured in the nonideal solutions of chemical and biological interest. This may be possible but attempts are just starting.

**Variational Approach.** I believe a variational approach designed to deal with strong interactions might be a useful alternative approach to the historical tradition[35-37], particularly if it can be modified to include interactions defined by simulations of molecular dynamics as seems possible (personal communication, Chun Liu). Ionic solutions in fact are a relatively simple complex fluid in some ways, because in the most important biological cases their microelements are hard spheres ($Na^+$, $K^+$, $Ca^{2+}$) or nearly hard spheres ($Cl^-$). Water can often be successfully described as a continuum, as it is in implicit solvent models of ionic solutions (also called 'the primitive model') and proteins.[10,17,18,32] The theory of complex fluids has dealt with systems with complex microelements: liquid crystals, polymeric fluids[38,39], colloids and suspensions[40] and electrorheological fluids[41]; magnetohydrodynamics systems[42]; systems with deformable electrolyte droplets that fission and fuse[28]; and suspensions of ellipsoids. The theory deals also interfacial properties of these complex mixtures, such as surface tension and the Marangoni effects of 'oil on water' and 'tears of wine'.[40]

It seems worthwhile to see how well the theory of complex fluids can deal with the key biological ions in water, $Na^+$, $K^+$, $Ca^{2+}$ and $Cl^-$. These ions are more (cations) or less (anions) hard spheres. They seem likely to have much less complex properties than the deformable charged droplets already treated by the theory of complex fluids.

But living solutions are not all that simple. Real extracellular solutions contain other components and the molecular detail of water can be important. Living solutions inside cells also contain proteins, nucleic acids, lipids, and organic ions (like free amino acids), that are complex microelements, that form the macromolecules of life. It will be interesting to see if the theory of complex fluids can be extended to them (in references[35-37] and forthcoming work involving membranes (Ryham, Liu, Eisenberg, and Cohen, personal communication) and tissue structures (Mori, Liu, and Eisenberg, personal communication).

We use[35] a theory of complex fluids based on the energy variational approach *EnVarA* of the mathematician Chun Liu who has actually provided the existence and uniqueness theorems needed to make this approach mathematics, as well as applying *EnVarA* to a variety of complex real systems[43-45]. We try to create a field theory of ionic solutions that uses only a few fixed parameters to calculate most properties in flow and in traditional thermodynamic equilibrium, both in bulk and in spatially complex domains like pores in channel proteins.

The Energy Variational Principle can be written as

$$
\overbrace{\frac{\delta E}{\delta \vec{x}}}^{\text{Conservative Force}} \quad = \quad \overbrace{\frac{1}{2}\frac{\delta \Delta}{\delta \vec{u}}}^{\text{Dissipative Force}}
\tag{1}
$$

The energy $E$ we use to describe finite size ions in a bulk solution is

$$E \ \text{Primitive Phase}; t \ =$$

$$
= \int_{\Omega} \left[ \underbrace{\underbrace{\frac{1}{2}\rho\left|\vec{u}_{IP}\right|^2}_{\text{Hydrodynamic Kinetic Energy}} + \underbrace{w(\rho)}_{\substack{\text{Hydrodynamic Potential Energy} \\ \text{Equation of State}}}}_{\text{Macroscopic (hydrodynamic)}} + \lambda\underbrace{\left[ \underbrace{\frac{1}{2}\varepsilon\left|\nabla\phi\right|^2}_{\text{Electrostatic}} + \underbrace{k_B T(c_n\log c_n + c_p\log c_p)}_{\text{Entropy}} + \underbrace{\psi\ \text{Solid Spheres}}_{\text{Finite Size Effect}} \right]}_{\text{Microscopic (atomic)}} \right] d\vec{x}
\tag{2}
$$

The dissipation $\Delta$ is not hard to derive but is too complex to present in detail because of the finite size effects. It is described in full in[35,36].

The variational principle *EnVarA* combines the maximum dissipation principle and least action principle into a force balance law that expands the conservative conservation laws to include dissipation, using the generalized forces in the variational formulation of mechanics (p. 19 of reference[46]; also[47]). This procedure is a modern reworking of Rayleigh's dissipation principle—eq. 26 of reference[48]—motivated by Onsager's treatment of dissipation[49,50]. *EnVarA* optimizes both the action functional (integral) of classical mechanics[51] and the dissipation functional[52]. The stationary point of the action is determined with respect to the trajectory of particles. The stationary point of the dissipation is determined with respect to rate functions (e.g., velocity). Both are written in Eulerian (laboratory) coordinates. These functionals can include entropy and dissipation as well as potential energy, and can be described in many forms on many scales from molecular dynamics calculations of atomic motion, to Monte Carlo MC simulations[44,53,54] to—more practically—continuum descriptions [55] of ions in water. We use a primitive model[10,17,18,25,56] of ions in an implicit solvent[57-61], adopting self-consistent treatments of electro-diffusion[62-65]—in which the charge on ions help create their own electric field—and introducing the repulsion energy of solid spheres[66,67], using the variational calculus to extend the primitive model to spatially complex, nonequilibrium time dependent situations, creating a field theory of ionic solutions.

Energy functional integrals and dissipation functional integrals are written from specific models of the assumed physics of a multi-component system, as did[27,28]. Components of the potential energy and dissipation functions are chosen so the variational procedure produces the drift diffusion equations of semiconductor physics[68]—called the Vlasov equations in plasma physics—or the similar biophysical Poisson Nernst Planck equations—named PNP by reference[62]—and used since then by many biophysicists[11,63,64,69-77] and physical chemists[65,78]. The energy of the repulsion of solid spheres can be included in the energy functionals in different ways using different forms for the interaction energy, giving similar but not identical results. It is included as Lennard-Jones spheres[79] giving (as their Euler-Lagrange equations) a generalization of PNP for solid ions. The energy of repulsion (for uncharged spheres) is included alternatively as in the density functional theory of fluids[80-82]. Boundary conditions tell how energy and matter flow into the system and from phase to phase and are described by a separate variational treatment of the 'interfacial' energy and dissipation. The resulting Euler Lagrange equations are the boundary value problems of our field theory of ionic solutions. They are derived by algebra and solved by mathematics—without additional physical approximations—in spatially complex domains, that perhaps produce flow of nonideal mixtures of ions in solution.

*EnVarA* does not produce a single boundary value problem or field equation for ionic solutions. Rather, it produces different field equations for different models (of correlations produced by screening or finite size, for example), to be checked by experiment. In the biological and chemical context *EnVarA* derives—it does NOT assume—systems of partial differential equations (i.e., field theories) of multiple interacting components and scales.

If a new component of energy (or dissipation) is added to a variational principle like *EnVarA*, the resulting Euler Lagrange equations—the field theory of electrolytes—change. The new field theory is derived by algebra and involves no further assumptions or parameters. The new field theory automatically includes all the interactions of the old and new components of the energy (and dissipation). This is an enormous advantage of variational principles and is probably the reason they are used so widely in physics. I am unaware of any other mathematical approach that forces field equations to be consistent with each other. The contrast with the usual approach to mixtures of ionic solutions, with their plethora of coupling coefficients, is striking. It is very difficult to determine those coupling coefficients, and even worse, the coupling coefficients are functions or functionals that depend on all the other parameters of the system, usually in an unknown way.

The variational principle can be applied to a primitive model of ionic solutions with a Lennard Jones treatment of excluded volume, and a selfconsistent computation of the electric field as described in detail in[35-37]. A regularized repulsive interaction potential is introduced as

$$\Psi_{i,j}(|\vec{x} - \vec{y}|) = \frac{\varepsilon_{i,j}(a_i + a_j)^{12}}{|\vec{x} - \vec{y}|^{12}} \tag{3}$$

for the $i^{th}$ and $j^{th}$ ions located at $\vec{x}$ and $\vec{y}$ with the radii $a_i$, $a_j$, respectively, where $\varepsilon_{i,j}$ is an empirically chosen energy constant,. Then the contribution of repulsive potential $\Psi$ to the total (free) energy is

$$E_{i,j}^{repulsion} = \tfrac{1}{2} \iint \Psi_{i,j}(|\vec{x} - \vec{y}|)c_i(\vec{x})c_j(\vec{y})\,d\vec{x}\,d\vec{y} \tag{4}$$

where $c_i$, $c_j$ are the densities of $i$ th, $j$ th ions, respectively.

For the sake of simplicity in this derivation, we consider a two-ion system with the charge densities, $c_n$, $c_p$. All derivations and programs have been written for a multiple ion system, with ions of any charge[35-37]. Then, the total repulsive energy is defined by

$$E^{repulsion} = \sum_{i,j=n,p} E_{i,j}^{repulsion} = \sum_{i,j=n,p} \tfrac{1}{2} \iint \Psi_{i,j}(|\vec{x} - \vec{y}|)c_i(\vec{x})c_j(\vec{y})\,d\vec{x}\,d\vec{y}. \tag{5}$$

Now we take a variational derivative with respect to each ion, $(\delta E^{repulsion}/\delta c_i) = 0$ to obtain the repulsive energy term and put it into the system of equations. This leads us to the following Nernst-Planck equations for the charge densities, $c_n$, $c_p$ :

$$\frac{\partial c_n}{\partial t} = \nabla \cdot \left[ D_n \left\{ \nabla c_n + \frac{c_n}{k_B T} \left( z_n e \nabla \phi - \int \frac{12\varepsilon_{n,n}(a_n + a_n)^{12}(\vec{x} - \vec{y})}{|\vec{x} - \vec{y}|^{14}} c_n(\vec{y})\,d\vec{y} \right. \right. \right.$$

$$\left. \left. \left. - \int \frac{6\varepsilon_{n,p}(a_n + a_p)^{12}(\vec{x} - \vec{y})}{|\vec{x} - \vec{y}|^{14}} c_p(\vec{y})\,d\vec{y} \right) \right\} \right], \tag{6}$$

$$\frac{\partial c_p}{\partial t} = \nabla \cdot \left[ D_p \left\{ \nabla c_p + \frac{c_p}{k_B T} \left( z_p e \nabla \phi - \int \frac{12\varepsilon_{p,p}(a_p + a_p)^{12}(\vec{x} - \vec{y})}{|\vec{x} - \vec{y}|^{14}} c_p(\vec{y})\,d\vec{y} \right. \right. \right.$$

$$\tag{7}$$

$$-\int\frac{6\varepsilon_{n,p}(a_n+a_p)^{12}(\vec{x}-\vec{y})}{|\vec{x}-\vec{y}|^{14}}c_n(\vec{y})\,d\vec{y}\Bigg)\Bigg\}\Bigg].$$

The details of the derivation of the repulsive terms in the chemical potentials are presented in [35-37] We now have the coupled system including finite size effects. We here call the system a modified PNP system. One advantage of the variational approach is the fact that the resulting system, the modified PNP, naturally satisfies the energy dissipation principle, the variational law eq. (1)

$$\frac{d}{dt}\int\Bigg\{k_BT\sum_{i=n,p}c_i\log c_i+\frac{1}{2}\Bigg(\rho_0+\sum_{i=n,p}z_iec_i\Bigg)\nabla\phi+\sum_{i,j=n,p}\frac{c_i}{2}\int\Psi_{i,j}c_j\,d\vec{y}\Bigg\}d\vec{x}$$

$$(8)$$

$$=-\int\Bigg\{\sum_{i=n,p}\frac{D_ic_i}{k_BT}\Bigg|k_BT\frac{\nabla c_i}{c_i}+z_ie\nabla\phi-\sum_{j=n,p}\nabla\int\tilde{\Psi}_{i,j}c_j\,d\vec{y}\Bigg|^2\Bigg\}d\vec{x}$$

where $\tilde{\Psi}_{i,j}=12\Psi_{i,j}$ for $i=j$, and $\tilde{\Psi}_{i,j}=6\Psi_{i,j}$ for $i\neq j$.

These variational principles derive field equations as we have seen that address and I believe will probably some day solve major problems in computational biology. The field theory *EnVarA* represents an ionic solution as a mixture of two fluids[83], a solvent water phase and an ionic phase. The ionic phase is a primitive model of ionic solutions. It is a compressible plasma made of charged solid (nearly hard) spheres. The ionic 'primitive phase' is itself a composite of two scales, a macroscopic compressible fluid and an atomic scale plasma of solid spheres in a frictional dielectric. Channel proteins are described by primitive ('reduced') models similar to those used to analyze the selectivity of calcium and sodium channels[84-86] and to guide the construction (using the techniques of molecular biology) of a real calcium channel protein in the laboratory[87,88]. Similar models predicted complex and subtle properties of the RyR channel before experiments were done in > 100 solutions and in 7 mutations, some drastic, removing nearly all permanent charge from the 'active site' of the channel (see references in[89,90]).

I believe a variational method is required to deal with real ionic solutions because ionic solutions are dominated by interactions. Ionic solutions do not resemble the ideal simple fluids of traditional theory and the interactions between their components are not two body, as assumed by the force fields of modern molecular dynamics. Indeed, **ions like Na⁺ and K⁺ have specific properties, and can be selected by biological systems, because they are non-ideal and have highly correlated behavior.** Screening and finite size effects produce the correlations more than anything else. Solvent effects enter (mostly) through the dielectric coefficient. Ionic solutions do not resemble a perfect gas[91] of non-interacting uncharged particles. Indeed, because of screening, the activity (which is a measure of the free energy) of an ionic solution is not an additive function as concentration is changed (Fig. 3.6 of reference[17]; Fig. 4.2.1 of reference [18]) and so does not easily fit some definitions of an extensive quantity (see p. 6 of the book of international standards for physical chemistry[92]).

Some correlations are included explicitly in our models as forces or energies that depend on the location of two particles. Other correlations are implicit and arise automatically as a mathematical consequence of optimizing the functionals *even if the models used in the functionals do not contain explicit interactions of components*. Kirchoff's current law (that implies perfect correlation in the flux of electrical charge[93]) arises this way as a consequence of Maxwell's equations[94] and does not need to be written separately.

Variational analysis is already an area of active research in modern mathematics. Our methods are also closely related to another exciting area of modern mathematical research, optimal control. Our *EnVarA* analysis produces 'optimal' estimates of the correlations that arise from those interactions (p. 42 of Gelfand and Fromin[95]; p. 11 of Biot[47]. Note criticism of Biot in[96]). All field equations arising from *EnVarA* optimize both the dissipation and the action integrals. Inadequate functionals can be corrected (to some extent) by adjusting effective parameters in the functionals.

Effective parameters are needed to deal with ions in electrolytes. Effective parameters are almost always used to describe complex interactions of ions in electrolyte solutions[26,97-101], e.g., the cross coupling Onsager coefficients[100-102] or Maxwell-Stefan coefficients[103]. *EnVarA* produces optimal estimates of these parameters, because the mathematics of variational analysis is almost identical to the mathematics of optimal control. Both use variational methods that can act on the same functionals. *EnVarA* becomes optimal control when the functionals are combined in a more general way than just adding them, e.g., by using Lagrange multipliers or more sophisticated techniques. Inverse methods[11,104,105] could be used to provide estimators of the parameters of *EnVarA* functionals with least variance or bias, or other desired characteristics. *EnVarA* gives the hope that fewer parameters can be used to describe a system than in models[56] and equations of state[19-21] of ionic solutions which involve many parameters. These parameters change with conditions and are really functions or even functionals of all the properties of the system. (It is important to understand that in general these coupling parameters need to depend on the type and concentration of all ions, not just the pair of ions that are coupled.)

Of course, the variational approach can only reveal correlations arising from the physics and components that the functional actually includes. Correlations arising from other components or physics need other models and will lead to other differential equations. For example, ionic interactions that arise from changes in the structure of water would be an example of 'other physics', requiring another model, if they could not be described comfortably by a change in the diffusion coefficient of an ion or a change in the dielectric constant of water. Numerical predictions of *EnVarA* will be relatively insensitive to the choice of description (of pairwise interactions, for example) because the variational process in general produces the 'optimal' result[47,95] for each version of the model. (This is an important practical advantage of the variational approach to optimal control: compare the success of the variational density functional theory of fluids[81,82] with the non-variational mean spherical approximation[17,18] that uses much the same physics.)

This variational approach can include energies of any type. It has in fact been used by Liu[44,53,54] to combine energies of reduced models and energies computed from simulations. It will be interesting to see how we can apply this approach to biological systems.

**Scaling in _EnVarA_ the Variational Approach**. The variational approach deals with issues of scaling in a very different way from direct simulations. _EnVarA_ has the great advantage of always being consistent. A model in _EnVarA_ is the statement of energies and dissipation in eq. (1). Once that model is chosen, the rest is algebra. The resulting Euler Lagrange equations form a well posed boundary value problem, a field theory of (usually) partial differential equations and boundary conditions that account for all the behavior of the system described by the energy and dissipation. The field theory is much more general than the thermodynamic and statistical mechanical ideas of equilibrium and state. It includes flow and interactions of components automatically. If two of the components of the energy (and/or dissipation) are on different scales, _EnVarA_ automatically produces Euler Lagrange equations that combine those scales selfconsistently. This is an enormous advantage compared to other multiscale methods. When dealing with interactions on one scale, and conservation laws on another, it is not at all easy to be sure that the resulting equations (corresponding to the field equations of _EnVarA_) are consistent, i.e., that the resulting equations satisfy the overriding constraints and conservation laws. When including the finite size of ions in classical theories of simple fluids, for example, it is very easy to use treatments that do not identically satisfy the equations of electrostatics. If the theory is meant to include electrodiffusion, and thereby extend to the nonequilibrium phenomena of life, it is very difficult to make the theory consistent with the special cases of diffusion of uncharged species (Fick's law), or the migration of charged species in systems without concentration gradients (Ohm's law), even if the theory ignores bulk flow and the complexities of hydrodynamic coupling.

_EnVarA_ deals with interactions automatically but it does not deal with multiscale issues nearly as well. We go through them one by one.

**Scaling in Space in _EnVarA_.** Spatial scaling and resolution are dealt with in _EnVarA_ without error if the models of energy and dissipation include all scales at perfect resolution. Of course, that never happens! What typically happens is that part of the system is known well at one scale, part at another, and parts of the system are left out. Typically, one part of the system must be resolved on one scale and the other on another. Applying _EnVarA_ to these situations is (reasonably) straightforward but the accuracy of the results can only be assessed after the fact by comparison with experiments. The basic approach is to write the energy and dissipation of each component of the model, of each scale, and combine them using Lagrange multiplier(s), or other penalty functions of optimal control. *EnVarA* guarantees that interactions will be dealt with correctly. *EnVarA* automatically deals with boundary conditions (once they are described with a model) and flow. These are important features not shared by many other methods. But *EnVarA* cannot deal with phenomena that are not present in the models of the energy and dissipation and these can be important. *EnVarA* (particularly when implemented numerically) may not be able to resolve steep phenomena and gradual phenomena well enough to estimate their interactions correctly. *EnVarA* will double count phenomena that are described in more than one component of a model. For example, if an equation of state is used to deal with the finite volume of ions (on the macroscopic scale) and Lennard Jones potentials are used to deal with the finite volume of ions (on the atomic scale), double counting can be expected. The Lagrange multipliers (or penalty functions of optimal control) and variational process minimizes the effect of the double counting (by choosing optimal parameters that minimize the functionals) but the residual effects may be significant. We are in unknown territory here. We know how to investigate but we do not know the results of the investigation.

**Scaling in time in *EnVarA***. Time dependence in *EnVarA* is produced by the dissipation function and so depends on the accuracy of the model of dissipation. It is obvious that the linear frictional model used in *EnVarA* (and in Rayleigh and Onsager's dissipation principles) is inadequate. Friction is not proportional to velocity in general. The consequences of the oversimplified model of dissipation are not known. At this stage, the time dependence computed with *EnVarA* seems to be that of the slowest 'time constant' of the system. Our working hypothesis is that the linear friction assumption produces a decent estimate of the (final) approach to equilibrium. It obviously cannot deal with complex time dependent phenomena that occur with complex friction. One way to deal with such phenomena is to include them as a separate component with a separate time scale and then to allow the variational process to do the matching between scales. This seems a different way of doing matching than in classical matched asymptotic expansions[106] but the literature has not been searched to verify that view. The issues involved in this approach are rather similar to those just discussed about spatial scales. Consistency is guaranteed between scales by the variational process, but double counting of some sort will occur. Investigation is needed and is underway.

One important characteristic of *EnVarA* arises from its time dependence and is both a curse and a blessing. The blessing is that *EnVarA* computes time dependence at all. The curse is that it must compute time dependence starting at time zero. Steady states only arise from transient computations. This property of the Euler Lagrange equations makes computation much less efficient. One must approach the steady state. One cannot just arrive there.

**Scaling of Parameters in *EnVarA*.** Parameters arise in *EnVarA* from the models of energy and dissipation and in general appear as parameters in the Euler Lagrange equations that specify the resulting field problem. Parameters are handled as well or as badly as they are in other partial differential equations. Analytically, parameters of any scale are handled 'perfectly', but numerical issues of stiffness and dynamic range can easily arise and be limiting. Each case must be studied as a separate numerical system because each case can have quite different qualitative behavior. The numerical schemes must be adapted to the qualitative behavior.

The very generality of the *EnVarA* approach causes considerable difficulty. The behavior of the system with all its interactions is often unknown in initial calculations. If reduced models with effective parameters are used (as they should be in early survey calculations), it is hard to know what 'region of phase space', i.e., what qualitative range of behaviors one is seeing. Dealing with an *EnVarA* calculation is much like a survey experiment in biology. You have to determine what is going on and you have to learn to simplify the calculation or experiment by choosing parameter ranges or setups in which the interesting phenomena dominate.

Computations of current flow through channels for example using *EnVarA* always produce charging phenomena at short times (because such must be present in any calculation that includes the electric field consistently), flow through the channel at intermediate times, and accumulation of ions outside the channel as the flow continues into long times. The charging phenomena and accumulation are peripheral to one's initial main interest in the channel itself, but the numerical procedures must deal with them correctly and efficiently. Experimental scientists may have taken years to learn to isolate the phenomena of interest. Numerical analysts using *EnVarA* face similar prospects.

A physical example may be helpful. Imagine trying to calculate the conductivity of a salt solution (or its 'dielectric constant' if you prefer an equilibrium property). In *EnVarA* one cannot assume good stirring or uniform temperature, unless one includes 'apparatus' (boundary conditions like stirrers or heat baths) that will do the stirring or supply the heat. The real system always has gradients of concentration and temperature, and *EnVarA* will always compute those because it is unable to calculate inconsistently even if we know the errors produced by the inconsistency are unimportant under the conditions of interest. Even worse, *EnVarA* computes these epi-phenomena in their full time dependent glory, even if we only want to know the steady state.

This power of *EnVarA* is again a blessing and a curse. It is a blessing because it forces the theorist to deal with phenomena well known in the laboratory (i.e., the difficulty of actually keeping solutions well stirred at constant temperature) but often not advertised in experimental papers. The curse is the difficulty of computation and the efforts needed to isolate important special cases.

Despite these difficulties, which are described here in vivid detail so we do not mislead the reader into thinking *EnVarA* is a magical solution for all problems, computations with *EnVarA* of real systems are possible. Many have been done in physical systems[28,38,39,42] and a substantial number have been done with some success in ionic solutions.[35-37] Once a system is understood, the difficulties just described are left behind, just as an experimental system goes quickly from 'impossible', to novel, to easy, to taken for granted after a few years of success. (Consider the history of single channel recording from 1975 to 1990 for example.)

**Scaling of the protein.** The above discussion does not deal with the multiscale issues of describing the protein, whether channel or enzyme. I do not know how to do that in a general way even for channels, where covalent bond changes and orbital delocalization are not involved, let alone for enzymes where covalent bond changes are what the system is all about. (See reference[107] for a discussion of 'Channels as Enzymes' and reference[4] for a discussion of channels as transistors.) Reduced models have been built in many ways, using quantum mechanics (references in[108]), reduced models with water detail (references in[109,110]), and reduced models with implicit models of water[67,75,111-115] and I apologize for the many references I have unknowingly omitted.

There seems to be no *a priori* way to choose between the different reduced models of channel proteins. I would use the fits to experimental data as the test for such models, although others prefer a more reductionist approach, arguing (understandably enough) that considerable structural detail is needed to deal with water and side chains of proteins. Each perspective emphasizes what the investigator can best do. My collaborators and I find that we can deal with nearly the whole range of experimental data on both calcium and sodium channels using a single model, with three parameters that never change value (the dielectric coefficients of protein and solution and the diameter of the channel) in a wide range of mixed solutions of different types and concentrations of ions of ions, using crystal radii of ions, even though calcium and sodium channels have very different properties[84-86,116]. This treatment uses grossly oversimplified models of the channel protein and its side chain but that simplification allows it to compute the Boltzmann distribution of structures of ions and side chains using Metropolis Monte Carlo methods. These methods show that the structure of the system changes significantly even dramatically as ions are changed in concentration or type. They show that the free energy of binding varies drastically as conditions are changed. Indeed, the model is used is a version of the self organized theory of proteins in which the fit of the ions to the active site and the fit of the active site to the ions is induced. The induced fit is determined as an output of the Monte Carlo simulations, as is the distribution of the fit. The model seems to work in a wide variety of conditions because it computes accurately, and guesses (with more luck than wisdom) the forces and energies actually used by biology to determine the selectivity of these channels to these ions.

**The role of biology.** The question arises: how can as complicated a system as a channel protein in a biological membrane surrounded by mixtures of ions be so simply described? The question is particularly vexing when one remembers that mixtures of ions in bulk solutions (without channel proteins) cannot be so simply described.

The reason seems to me biological and evolutionary.

Biological systems are not general physical systems. Biological systems have been built by evolution to have definite functions. Evolution acts by mutating genes and genes make proteins. Proteins are coded amino acid by amino acid, and mutations change individual amino acids. It seems obvious that a system like this will discover 'controls' which produce useful functions. (Useful functions are those that allow their host organism to survive natural selection.) Individual amino acids will control individual functions in such a system. These thoughts are hardly rigorous, but they provide motivation to accept the experimental fact that individual amino acids do control function in many important cases. In many cases a few amino acids or a particular structural domain of a protein controls overall function.

Viewed from an engineering perspective, this biological simplicity is not a surprise. Devices are built so they can be controlled. The control of a device is often far more important than its efficiency. An easy way to ensure robust control is to put that control in a separate system distinct from the rest of the device. Evolution seems to use that approach. Evolution has found ways to use only a few amino acids to control biological function. The physics does not force this. Evolution has.

Reduced models that describe so many properties of dissimilar calcium and sodium channels with so few parameters should be viewed as the expression of evolution. These models must be describing the energies evolution has used to produce these functions. It seems likely that energies are correct in a model with just two parameters that fits data from two different types of channels, with quite different properties, in many solutions and concentrations, using crystal radii of ions, and parameter values that are the same under all conditions and in both channel types. It seems likely that evolution has chosen to create this kind of selectivity (for salts) in this kind of channel ($Ca^{2+}$ and $Na^+$ channels) in this way and the investigators of these channels could study these energies because they were relatively simple to compute (although it still took many years and many papers and methods).

There is no guarantee, however, that these energies will be the only energies used to determine the selectivity of other types of ions or other systems (e.g., the zinc finger binding system so well studied experimentally[117,118]). However, it seems likely that these energies will be involved, along with others.

In that case, it seems safe to say that simulations that try to deal with selectivity or biological function using a single free energy of binding, that does not vary with ion type or concentration in the baths, will be inadequate, unable to deal with the essentials of biological function.

The binding data we have computed is an incomplete description of biological reality in an important way. The Monte Carlo method is constrained to equilibrium, in fact to zero concentration and zero electrical potential gradients in the way we do it. Biology does not occur under these conditions. Our simple model of binding has been extended to nonequilibrium systems using a hybrid of the density functional theory of nonelectrolytes (of Rosenfeld[119-123]; note this has nothing to do with the density functional theory of electrons) and the Poisson Nernst Planck theory of ionic solutions, named PNP in reference[62] and used by many workers in chemistry[65] and biophysics[109,124-128] since then. The resulting DFT-PNP theory has been applied by Gillespie and co-workers with some considerable success to the Ryanodine Receptor channel of the heart. They have shown excellent agreement between experiment and data and have predicted experimental results before the experiments were performed.[82,89,122,129] But there are problems because electrostatics are added to DFT in an imaginative but ad hoc manner[82,122,129] that suffers fundamental difficulties. In particular,

(a) PNP-DFT does not satisfy Gauss' law or sum rules, as it should.

(b) PNP-DFT is not derived from a general variational principle and so is ad hoc and incomplete as well as imaginative and powerful. PNP-DFT omits the important electrophoretic, relaxation, hydrodynamic, and osmotic components of current, found in experiments and in all theories since the 1930's work of Onsager & Fuoss[26,130].

In my opinion, PNP-DFT is a useful beginning but *EnVarA* has a greater future because it is based on fundamental principles, satisfies sum rules, and yields all interactions of all species. *EnVarA* is a superset of DFT (of neutral species) and a super set of PNP-DFT. *EnVarA* should be able to do much more. It has not done that yet, but many investigators are trying.

**Conclusions**. What can we say then in general about the computation of biological systems?

We can say that

1) simulation in atomic detail is unlikely to succeed because of the scaling issues shown in Table 1.

2) simulations must be calibrated against experimental data in realistic mixed solutions because those are the only conditions in which living systems function.

3) reduced models are needed because nothing else is likely to deal with the scaling issues.

4) reduced models of ionic solutions should be based on a mathematics of interacting systems, a variational principle like *EnVarA,* because that automatically deals consistently with multiple interacting components. It deals with multifaceted interactions without introducing many underdetermined coupling parameters and coefficients.

5) reduced models of channel proteins may take on many forms, but they must deal with the range of ionic conditions in which the channels actually work, even if these are mixtures of ions uncomfortable to compute because so many ions are so important over such a wide range of concentrations.

6) a particular reduced model of calcium and sodium channels has been surprisingly successful. Similar models should be tried in other systems, hoping that their simplicity will at least point in the right direction to help uncover relevant computable complexity.

I look forward to the next Festschrift for Mark Ratner and hope my colleagues and I can show him then how we have dealt with multiscale issues in a variety of new systems.

**Acknowledgement**

It is a pleasure to thank Mark Ratner for his encouragement and help at a critical point in my transformation from experimental biophysicist to striving biophysical chemist. Many of my publications would not have been written if he had not introduced me to Zeev Schuss, Ron Elber, and so much of modern physical chemistry.

This paper reports the work of a wide group of collaborators without whom none of it would have happened. I am grateful beyond words for the opportunity (and joy) of working with them. The paper was edited with great skill by Ardyth Eisenberg to whom I owe very much beyond that. This work was supported in part by NIH grant GM076013.

**Table 1**

| Computational Scale | Biological Scale | Ratio |
|---|---|---|
| **Time** $10^{-16}$ sec<br>*Vibrations of Bonds* | $10^{-5}$ sec<br>*Action Potential* | $10^{11}$ |
| **Space** $10^{-11}$ m<br>*Side Chains of Protein* | $10^{-4}$ m<br>Large Cell | $10^{7}$ |
| **Volume** | | $10^{21}$ |
| **Spatial Resolution** | | $10^{9}$ |
| **Solute Concentration** | $10^{-11}$ to $2\times10^{1}$ M | $10^{12}$ |

Scaling restrictions implied by the long range electric field are not clear because the accuracy of the Ewald sum treatment of periodic boundary conditions is not clear. See text.

**References**

(1)     Moore, G. E. "Lithography and the future of Moore's law", 1995, Santa Clara, CA, USA.

(2)     Alberts, B.; Bray, D.; Lewis, J.; Raff, M.; Roberts, K.; Watson, J. D. *Molecular Biology of the Cell*, Third ed.; Garland: New York, 1994.

(3)     Weiss, T. F. *Cellular Biophysics*; MIT Press: Cambridge MA USA, 1996; Vol. 1 and 2.

(4)     Eisenberg, B. *http://arxiv.org/* **2005**, *q-bio.BM*, arXiv:q.

(5)     Dixon, M.; Webb, E. C. *Enzymes*; Academic Press: New York, 1979.

(6)     Hille, B. *Ionic Channels of Excitable Membranes*, 3rd ed.; Sinauer Associates Inc.: Sunderland, 2001.

(7)     Sakmann, B.; Neher, E. *Single Channel Recording.*, Second ed.; Plenum: New York, 1995.

(8)     Henderson, J. R. Statistical Mechanical Sum Rules. In *Fundamentals of Inhomogeneous Fluids*; Henderson, D., Ed.; Marcel Dekker: New York, 1992; pp 23.

(9)     Barthel, J.; Buchner, R.; Münsterer, M. *Electrolyte Data Collection Vol. 12, Part 2: Dielectric Properties of Water and Aqueous Electrolyte Solutions*; DECHEMA: Frankfurt am Main, 1995.

(10)    Barthel, J.; Krienke, H.; Kunz, W. *Physical Chemistry of Electrolyte Solutions: Modern Aspects*; Springer: New York, 1998.

(11)    Burger, M.; Eisenberg, R. S.; Engl, H. *SIAM J Applied Math* **2007**, *67*, 960.

(12)    Damocles. *Damocles Web Site, IBM Research*. In *http://www.research.ibm.com/DAMOCLES/home.html*, 2007.

(13)    Markowich, P. A.; Ringhofer, C. A.; Schmeiser, C. *Semiconductor Equations*; Springer-Verlag: New York, 1990.

(14)    Selberherr, S. *Analysis and Simulation of Semiconductor Devices*; Springer-Verlag: New York, 1984.

(15)    Tanford, C.; Reynolds, J. *Nature's Robots: A History of Proteins*; Oxford: New York, 2001.

(16)    Zhang, C.; Raugei, S.; Eisenberg, B.; Carloni, P. *Journal of Chemical Theory and Computation* **2010**, *6*, 2167.

(17)    Fawcett, W. R. *Liquids, Solutions, and Interfaces: From Classical Macroscopic Descriptions to Modern Microscopic Details*; Oxford University Press: New York, 2004.

(18)    Lee, L. L. *Molecular Thermodynamics of Electrolyte Solutions*; World Scientific Singapore, 2008.

(19)    Sengers, J. V.; Kayser, R. F.; Peters, C. J.; White, H. J., Jr. *Equations of State for Fluids and Fluid Mixtures (Experimental Thermodynamics)* Elsevier: New York, 2000.

(20)    Lin, Y.; Thomen, K.; Hemptinne, J.-C. d. *American Institute of Chemical Engineers AICHE Journal* **2007**, *53*, 989.

(21)    Jacobsen, R. T.; Penoncello, S. G.; Lemmon, E. W.; Span, R. Multiparameter Equations of State. In *Equations of State for Fluids and Fluid Mixtures*; Sengers, J. V., Kayser, R. F., Peters, C. J., White, H. J., Jr., Eds.; Elsevier: New York, 2000; pp 849.

(22)     Hansen, J.-P.; McDonald, I. R. *Theory of Simple Liquids*, Third Edition ed.; Academic Press: New York, 2006.

(23)     Rice, S. A.; Gray, P. *Statistical Mechanics of Simple Fluids*; Interscience (Wiley): New York, 1965.

(24)     Barker, J.; Henderson, D. *Reviews of Modern Physics* **1976**, *48*, 587.

(25)     Durand-Vidal, S.; Simonin, J.-P.; Turq, P. *Electrolytes at Interfaces*; Kluwer: Boston, 2000.

(26)     Justice, J.-C. Conductance of Electrolyte Solutions. In *Comprehensive Treatise of Electrochemistry Volume 5 Thermondynbamic and Transport Properties of Aqueous and Molten Electrolytes*; Conway, B. E., Bockris, J. O. M., Yaeger, E., Eds.; Plenum: New York, 1983; pp 223.

(27)     Yue, P.; Feng, J. J.; Liu, C.; Shen, J. *Journal of Fluid Mechanics* **2004**, *515*, 293.

(28)     Ryham, R.; Liu, C.; Zikatanov, L. *Discrete and Continuous Dynamical Systems-Series B* **2007**, *8*, 649.

(29)     Zhang, J.; Gong, X.; Liu, C.; Wen, W.; Sheng, P. *Physical Review Letters* **2008**, *101*, 194503.

(30)     Sheng, P.; Zhang, J.; Liu, C. *Progress of Theoretical Physics Supplement No. 175* **2008**, 131.

(31)     Singer, A.; Schuss, Z.; Eisenberg, R. S. *Journal of Statistical Physics* **2005**, *119*, 1397.

(32)     Fraenkel, D. *Molecular Physics* **2010**, *108*, 1435

(33)     Pitzer, K. S. *Activity Coefficients in Electrolyte Solutions*; CRC Press: Boca Raton FL USA, 1991.

(34)     Post, D. E.; Votta, L. G. *Physics Today* **2005**, *58*, 35.

(35)     Eisenberg, B.; Hyon, Y.; Liu, C. *Journal of Chemical Physics* **2010**, *133*, 104104

(36)     Hyon, Y.; Eisenberg, B.; Liu, C. *Communications in Mathematical Sciences (in the press) also available as preprint# 2318 (IMA, University of Minnesota, Minneapolis)* *http://www.ima.umn.edu/preprints/jun2010/jun2010.html* **2010**.

(37)     Eisenberg, B. *Advances in Chemical Physics (in the press)* **2010**, *also available at http:\\arix.org as Paper arXiv 1009.1786v1*

(38)     Bird, R. B.; Armstrong, R. C.; Hassager, O. *Dynamics of Polymeric Fluids, Fluid Mechanics*; Wiley: New York, 1977; Vol. Volume 1.

(39)     Bird, R. B.; Hassager, O.; Armstrong, R. C.; Curtiss, C. F. *Dynamics of Polymeric Fluids, Kinetic Theory* Wiley: New York, 1977; Vol. Volume 2.

(40)     Yue, P.; Feng, J. J.; Liu, C.; Shen, J. *Journal of Fluid Mechanics* **2005**, *540*, 427.

(41)     Cheng, Y.; Chang, C. E.; Yu, Z.; Zhang, Y.; Sun, M.; Leyh, T. S.; Holst, M. J.; McCammon, J. A. *Biophys J* **2008**, *95*, 4659.

(42)     Liu, C.; Liu, H. *SIAM Journal of Applied Mathematics* **2008**, *68*, 1304.

(43)     Lei, Z.; Liu, C.; Zhou, Y. *Commun. Math. Sci.* **2007**, *5*, 595.

(44)     Du, Q.; Hyon, Y.; Liu, C. *Journal of Multiscale Modeling and Simulation* **2008**, *2*, 978.

(45)     Hyon, Y.; Kwak, D. Y.; Liu, C. *Discrete and Continuous Dynamical Systems (DCDS-A)* **2010** *26*, 1291

(46)     Goldstein, H. *Classical Mechanics, Second Edition*, 2nd ed.; Addison Wesley: Reading, MA, 1980.

(47)     Biot, M. A. *Variational Principles in Heat Transfer: A Unified Lagrangian Analysis of Dissipative Phenomena*; Oxford University Press: New York, 1970.

(48)    Rayleigh, L., previously John William Strutt. *Proceedings of the London Mathematical Society* **1873**, *IV*, 357.

(49)    Onsager, L. *Physical Review* **1931**, *37*, 405.

(50)    Onsager, L. *Physical Review* **1931**, *38*, 2265.

(51)    Arnold, V. I. *Mathematical Methods of Classical Mechanics, 2nd Editiona*, 2nd ed.; Springer: New York, 1997.

(52)    Landau, L. D.; Lifshitz, E. M. *Course of Theoretical Physics, Volume 5: Statistical Physics.*, 3rd edition ed.; Butterworth Heinemann: London, 1996; Vol. Volume 5.

(53)    Du, Q.; Liu, C.; Yu, P. *Multiscale Modeling & Simulation* **2005**, *4*, 709.

(54)    Yu, P.; Du, Q.; Liu, C. *Multiscale Modeling & Simulation* **2005**, *3*, 895.

(55)    Lin, F.-H.; Liu, C.; Zhang, P. *Communications on Pure and Applied Mathematics* **2005**, *58*, 1437.

(56)    Pitzer, K. S. *Thermodynamics*, 3rd ed.; McGraw Hill: New York, 1995.

(57)    Warshel, A.; Russell, S. T. *Quarterly Review of Biophysics* **1984**, *17*, 283.

(58)    Gilson, M. K.; Honig, B. *Biopolymers* **1985**, *25*, 2097.

(59)    Davis, M. E.; McCammon, J. A. *Chem. Rev.* **1990**, *90*, 509.

(60)    Antosiewicz, J.; McCammon, J. A.; Gilson, M. K. *Biochemistry* **1996**, *35*, 7819.

(61)    Roux, B. Implicit solvent models. In *Computational Biophysics*; Becker, O., MacKerrel, A. D., B., R., Watanabe, M., Eds.; Marcel Dekker Inc: New York, 2001; pp p. 133.

(62)    Eisenberg, R.; Chen, D. *Biophysical Journal* **1993**, *64*, A22.

(63)    Eisenberg, R. S. *J. Membrane Biol.* **1996**, *150*, 1.

(64)    Eisenberg, R. S. Atomic Biology, Electrostatics and Ionic Channels. In *New Developments and Theoretical Studies of Proteins*; Elber, R., Ed.; World Scientific: Philadelphia, 1996; Vol. 7; pp 269.

(65)    Bazant, M. Z.; Thornton, K.; Ajdari, A. *Physical Review E* **2004**, *70*, 021506.

(66)    Nonner, W.; Catacuzzeno, L.; Eisenberg, B. *Biophysical Journal* **2000**, *79*, 1976.

(67)    Eisenberg, B. *Biophysical Chemistry* **2003**, *100*, 507

(68)    Jerome, J. W. *Analysis of Charge Transport. Mathematical Theory and Approximation of Semiconductor Models*; Springer-Verlag: New York, 1995.

(69)    Eisenberg, B. *Physics Today* **2006**, *59*, 12.

(70)    Schuss, Z.; Nadler, B.; Eisenberg, R. S. *Phys Rev E Stat Nonlin Soft Matter Phys* **2001**, *64*, 036116.

(71)    Hollerbach, U.; Chen, D. P.; Busath, D. D.; Eisenberg, B. *Langmuir* **2000**, *16*, 5509.

(72)    Eisenberg, B. *Permeation as a Diffusion Process*; http://www.biophysics.org/btol/channel.html#5, 2000.

(73)    Hollerbach, U.; Chen, D.; Nonner, W.; Eisenberg, B. *Biophysical Journal* **1999**, *76*, A205.

(74)    Eisenberg, R. S. *Journal of Membrane Biology* **1999**, *171*, 1.

(75)    Mamonov, A. B.; Coalson, R. D.; Nitzan, A.; Kurnikova, M. G. *Biophys J* **2003**, *84*, 3646.

(76)    Corry, B.; Kuyucak, S.; Chung, S. H. *Biophys J* **2003**, *84*, 3594.

(77)    Im, W.; Roux, B. *Journal of Molecular Biology* **2002**, *319*, 1177.

(78)    Newman, J.; Thomas-Alyea, K. E. *Electrochemical Systems*, 3rd ed.; Wiley-Interscience: New York, 2004.

(79)    Lin, F.-H.; Liu, C.; Zhang, P. *Communications on Pure and Applied Mathematics* **2007**, *60*, 838.

(80)    Davis, H. T. *Statistical Mechanics of Phases, Interfaces, and Thin Films*; Wiley-VCH: New York, 1996.

(81)    Roth, R.; Evans, R.; Lang, A.; Kahl, G. *J. Phys.: Condens. Matter* **2002**, *14*, 12063.

(82)    Gillespie, D.; Nonner, W.; Eisenberg, R. S. *Journal of Physics (Condensed Matter)* **2002**, *14*, 12129.

(83)    Liu, C.; Shen, J. *Physica D: Nonlinear Phenomena* **2003**, *179*, 211.

(84)    Eisenberg, B. *Institute of Mathematics and its Applications* **2009**, *IMA University of Minnesota* http://www.ima.umn.edu/2008.

(85)    Boda, D.; Valisko, M.; Henderson, D.; Eisenberg, B.; Gillespie, D.; Nonner, W. *J. Gen. Physiol.* **2009**, *133*, 497.

(86)    Boda, D.; Nonner, W.; Valisko, M.; Henderson, D.; Eisenberg, B.; Gillespie, D. *Biophys. J.* **2007**, *93*, 1960.

(87)    Miedema, H.; Meter-Arkema, A.; Wierenga, J.; Tang, J.; Eisenberg, B.; Nonner, W.; Hektor, H.; Gillespie, D.; Meijberg, W. *Biophys J* **2004**, *87*, 3137.

(88)    Vrouenraets, M.; Wierenga, J.; Meijberg, W.; Miedema, H. *Biophys J* **2006**, *90*, 1202.

(89)    Gillespie, D.; Xu, L.; Wang, Y.; Meissner, G. *Journal of Physical Chemistry* **2005**, *109*, 15598.

(90)    Gillespie, D. *Biophys J* **2008**, *94*, 1169.

(91)    Rowlinson, J. S. *The Perfect Gas*; Macmillan: New York, 1963.

(92)    Cohen, E. R.; Cvitas, T.; Frey, J.; Holmstrom, B.; Kuchitsu, K.; Marquardt, R.; Mills, I.; Pavese, F.; Quack, M.; Stohner, J.; Strauss, H. L.; Takami, M.; Thor, A. J. *Quantities, Units and Symbols in Physical Chemistry*, Third Edition ed.; Royal Society of Chemistry Publishing: Cambridge, UK, 2007.

(93)    Heras, J. A. *American journal of physics* **2008**, *76*, 101.

(94)    Nonner, W.; Peyser, A.; Gillespie, D.; Eisenberg, B. *Biophys J* **2004**, *87*, 3716.

(95)    Gelfand, I. M.; Fromin, S. V. *Calculus of Variations*; Dover: New York, 1963.

(96)    Finlayson, B. A. *The method of weighted residuals and variational principles: with application in fluid mechanics, heat and mass transfer*; Academic Press: New York, 1972.

(97)    Roger, G. l. M.; Durand-Vidal, S.; Bernard, O.; Turq, P. *The Journal of Physical Chemistry B* **2009**, *113*, 8670.

(98)    Dufreche, J. F.; Bernard, O.; Turq, P.; Mukherjee, A.; Bagchi, B. *Phys Rev Lett* **2002**, *88*, 095902.

(99)    Durand-Vidal, S.; Turq, P.; Bernard, O.; Treiner, C.; Blum, L. *Physica A* **1996**, *231*, 123.

(100)   DeGroot, S. R.; Mazur, P. *Non-Equilibrium Thermodynamics.*; North-Holland Publishing Co.: Amsterdam, 1962.

(101)   Katchalsky, A.; Curran, P. F. *Nonequilibrium Thermodynamics*; Harvard: Cambridge, MA, 1965.

(102)   DeGroot, S. R. *Thermodynamics of Irreversible Processes*; North-Holland: Amsterdam, 1961.

(103)   Taylor, R.; Krishna, R. *Multicomponent Mass Transfer*; Wiley: New York, 1993.

(104)   Engl, H. W.; Hanke, M.; Neubauer, A. *Regularization of Inverse Problems* Kluwer: Dordrecht, The Netherlands, 2000.

(105)   Kaipio, J.; Somersalo, E. *Statistical and Computational Inverse Problems* Springer: New York, 2005.

(106)   Kevorkian, J.; Cole, J. D. *Multiple Scale and Singular Perturbation Methods*; Springer-Verlag: New York, 1996.

(107)   Eisenberg, R. S. *Journal of Membrane Biology* **1990**, *115*, 1.

(108)   Varma, S.; Sabo, D.; Rempe, S. B. *J Mol Biol* **2008**, *376*, 13.

(109)   Roux, B. *Biophys J* **2010**, *98*, 2877.

(110)   Bostick, D. L.; Brooks, C. L., 3rd. *Biophys J* **2009**, *96*, 4470.

(111)   Nonner, W.; Chen, D. P.; Eisenberg, B. *Biophysical Journal* **1998**, *74*, 2327.

(112)   Wang, Y.; Xu, L.; Pasek, D.; Gillespie, D.; Meissner, G. *Biophysical Journal* **2005**, *89*, 256.

(113)   Allen, T. W.; Kuyucak, S.; Chung, S. H. *Biophys J* **1999**, *77*, 2502.

(114)   Corry, B.; Kuyucak, S.; Chung, S. H. *J Gen Physiol* **1999**, *114*, 597.

(115)   Corry, B.; Chung, S.-H. *European Biophysics Journal* **2005**, *34*, 208.

(116)   Boda, D.; Nonner, W.; Henderson, D.; Eisenberg, B.; Gillespie, D. *Biophys. J.* **2008**, *94*, 3486.

(117)   Berg, J. M. *J Biol Chem* **1990**, *265*, 6513.

(118)   Shi, Y.; Berg, J. M. *Chem Biol* **1995**, *2*, 83.

(119)   Rosenfeld, Y. Geometrically based density-functional theory for confined fluids of asymmetric ("complex") molecules. In *Chemical Applications of Density-Functional Theory*; Laird, B. B., Ross, R. B., Ziegler, T., Eds.; American Chemical Society: Washington, D.C., 1996; Vol. 629; pp 198.

(120)   Evans, R. Density Functionals in the Theory of Nonuniform Fluids. In *Fundamentals of Inhomogeneous Fluids*; Henderson, D., Ed.; Marcel Dekker: New York, 1992; pp 606.

(121)   Goulding, D.; Melchionna, S.; Hansen, J.-P. *Phys Chem Chem Physics* **2001**, *3*, 1644.

(122)   Gillespie, D.; Nonner, W.; Eisenberg, R. S. *Physical Review E* **2003**, *68*, 0313503.

(123)   Roth, R. *Journal of Physics: Condensed Matter* **2010**, *22*, 063102.

(124)   Fritsch, N.; Pouquet, O.; Roux, B.; Abdelmoumen, Y.; Janvier, G. *Ann Fr Anesth Reanim* **2010**, *29*, 45.

(125)   Roux, B.; Yu, H. *J Chem Phys* **2010**, *132*, 234101.

(126)   Roux, B. *J Gen Physiol* **2010**, *135*, 547.

(127)   Egwolf, B.; Luo, Y.; Walters, D. E.; Roux, B. *J Phys Chem B* **2010**, *114*, 2901.

(128)   Chakrapani, S.; Sompornpisut, P.; Intharathep, P.; Roux, B.; Perozo, E. *Proc Natl Acad Sci U S A* **2010**.

(129)   Gillespie, D.; Valisko, M.; Boda, D. *Journal of Physics: Condensed Matter* **2005**, *17*, 6609.

(130)   Fuoss, R. M.; Onsager, L. *Proc Natl Acad Sci U S A* **1955**, *41*, 274.